\documentclass[twocolumn, aps, pra, 10pt, a4paper]{revtex4-2}
\usepackage{amssymb}
\usepackage{amsmath}
\usepackage{amsfonts}
\usepackage{graphicx}
\usepackage{xcolor}
\usepackage{orcidlink}
\hypersetup{
	colorlinks=true,
	linkcolor=blue,
	filecolor=magenta,
	urlcolor=cyan,
	citecolor=red
}
\usepackage{hyperref}
\begin{document}
\title{Relaxation in Polymer Networks under Uniaxial Extension and Biaxial Compression}
\author{Volker Kraus$^1$\orcidlink{0000-0002-5577-8077}}
\author{Wolfgang Hamm$^1$}
\author{Miklos Zrinyi$^2$}
\affiliation{$^1$Institut für Exp. Physik, Universität Ulm, Albert--Einstein--Allee 11, 89077 Ulm, Germany}
\affiliation{{$^2$Institut für Biophysik und Strahlenbiologie -- Semmelweis Universität},
             {Üllői út 26},
             {H-1085 Budapest},
             {Hungary}}
\begin{abstract}
Predicting the time- and temperature-dependent behavior of polymer networks under complex loading is essential for the design of advanced elastomeric materials. Many practical applications involve combinations of deformation modes — such as uniaxial extension and biaxial compression — yet a unified description of their mechanical response remains challenging.
In this study, we apply a consistent theoretical framework to describe both uniaxial and biaxial deformation modes, using the same constitutive formalism based on van der Waals network theory. The time dependence of the material response in both cases is governed by a 
substance-specific relaxation spectrum, introduced through irreversible thermodynamics as a linear coupling to the quasi-static reference state of the permanent network. The temperature dependence of the relaxation times is well described by the Williams–Landel–Ferry (WLF) equation in the high-temperature or low strain-rate regime, demonstrating that the same physical mechanisms underlie time-dependent behavior across different loading geometries.
Experimental results are presented for cross-linked poly(methyl methacrylate) (PMMA) and polyvinyl acetate (PVAc), validating the theoretical model across both materials and deformation modes.
\end{abstract}
\maketitle
\section{Introduction}
\label{sec1}
Far above the glass transition temperature $T_G$, stress strain-curves of rubbers under quasi static conditions have been described up to large deformations \cite{flory_principles_1953,treloar_physics_1967}. 
Since then quite a number of constitutive models have been developed to describe the strain-energy functions for rubber and rubber-like materials. 
An overview of different constitutive models including the van der Waals model can be found in \cite{he_comparative_2022}. We prefer the van der Waals network theory because its strain–energy function involves only a small number of parameters, each of which has a clear physical interpretation. \cite{vilgis_equilibrium_1986,zrinyi_decisive_1989}.
It might be worth mentioning that recently in the area of biopolymers, the van der Waals model was successfully used to describe the behavior of lung tissue deformation under load \cite{andrikakou_behaviour_2016, naumann_mechanical_2022}.
\\
The investigation of time-dependent phenomena in permanent polymer networks subjected to large deformations beyond simple uniaxial extension serves as a foundation for consolidating and validating the theoretical framework. In particular, it substantiates the application of the van der Waals network theory in conjunction with the thermodynamics of irreversible processes to describe the viscoelastic behavior of rubber-like materials above the glass transition temperature. As demonstrated in this work, this approach accurately captures not only uniaxial but also the biaxial deformation mode. Moreover, the formalism can be further extended to cover the temperature regime near and above the glass transition, significantly broadening its range of applicability.
\\
Away from the equilibrium state, a formalism has been defined for the treatment of the stress-strain relationships at different temperatures and at different strain rates.
Hysteresis effects of rubbers during stress--strain cycles at different temperatures and under different strain rates are well understood \cite{ambacher_relaxation_1989,enderle_irreversible_1984}
based on a formalism of the thermodynamics of irreversible processes \cite{onsager_reciprocal_1931,onsager_reciprocal_1931-1}. The Gibbs function of a van der Waals network, which is used to describe the equilibrium stress--strain curve,
is extended by an adequate set of hidden variables \cite{degroot_non-equilibrium_1962,lebon_understanding_2008, bikkin_non-equilibrium_2021,di_vita_non-equilibrium_2022}.
These hidden variables represent elementary relaxation modes, which are, in terms of a relaxation mode coupling model \cite{kraus_relaxation_1994}, coupled to the global level of the network (equilibrium state) in a linear and scalar way. 
\\
The successful agreement with experimental data confirms that the stress–strain response at large deformations, over a range of temperatures and strain rates, can be fully characterized by the shear relaxation spectrum obtained from small-strain experiments \cite{kraus_relaxation_1992}.
This can be explained by a linear response, which is characterized by a system--typical strain--independent relaxation time spectrum, which was measured in a dynamic-mechanical relaxation experiment \cite{wrana_introduction_2009}. 
When approaching the glass transition, stress--strain cycles display macroscopically increasing nonlinear features especially in the range of small strains, like yielding. Up  to the glass transition our theory provides however good agreement for both uniaxial and biaxial deformation modes. 
\section{Materials and Methods}
\subsection{Equilibrium Approach}
Based on the van der Waals network \cite{kilian_thermomechanik_1979,kilian_molecular_1980,kilian_equation_1981,eisele_new_1981}, the van der Waals strain energy function $W$ as a function of the strain $\lambda$ reads \cite{enderle_irreversible_1984}
\begin{equation}\label{vdw1}
W(\lambda) = -G \left\lbrace 2\,\Phi_m 
[ \ln (1 - \eta) + \eta ]+\frac{2}{3} 
a\,\Phi^{3/2} \right\rbrace
\end{equation}
\begin{equation}\label{vdw1.1}
 = G\,w(\lambda)
\end{equation}
with the deformation function $\Phi(\lambda)=\frac{1}{2}({\lambda^2}+\frac{2}{\lambda}-3)$ in the mode of simple extension, 
$\Phi_m=\Phi(\lambda_m)$ and $\eta=\sqrt{\Phi/ \Phi_m}$.
According to \cite{treloar_physics_1967} the mode of uniaxial compression is formally identical to simple extension, so we may apply (\ref{vdw1}) directly, which fits e.g. for PMMA as shown in Fig.\ref{fig.3}.\\
The first of the strain independent van der Waals parameters, the maximum strain $\lambda_m$, describes the maximum chain extensibility in networks with finite chain length. The second parameter $a$ is a phenomenological parameter which characterizes global interactions between the fluctuating cross links of the network chains. This results in a reduction of the macroscopic force, which is to be measured \cite{vilgis_equilibrium_1986}.
The shear modulus of a permanent network \cite{kilian_rubber_1986} reads
\begin{eqnarray}\label{vdw2}
G&=&\frac{\rho\,R\,T}{M_u\,\lambda^{2}_{m}} \quad ,
\end{eqnarray} 
where $R$ is the gas constant, $T$ the absolute temperature,
$M_u$ the molecular weight of the stretching invariant unit 
and
$\rho$ the density of the polymer.
Eqn. (\ref{vdw1}) yields
the van der Waals equation of state for the stress $f$ \cite{enderle_irreversible_1984}
\begin{equation}\label{vdw4}
f(\lambda)=G\,D(\lambda)
\left\{ \frac{1}{1-\eta}-a\,{\Phi}^{\frac{1}{2}}(\lambda)\right\}
\end{equation}
with $D(\lambda)=\displaystyle\frac{\partial}{\partial\lambda}\Phi(\lambda)$.
Equation (\ref{vdw4}) fits well to experimental data in uniaxial and biaxial mode as shown in Fig.\ref{fig.2}. The experimental set-up for biaxial compression is described in \cite{hamm_uniaxiale_1993}.
\begin{figure}[ht]
\includegraphics[scale=0.365]{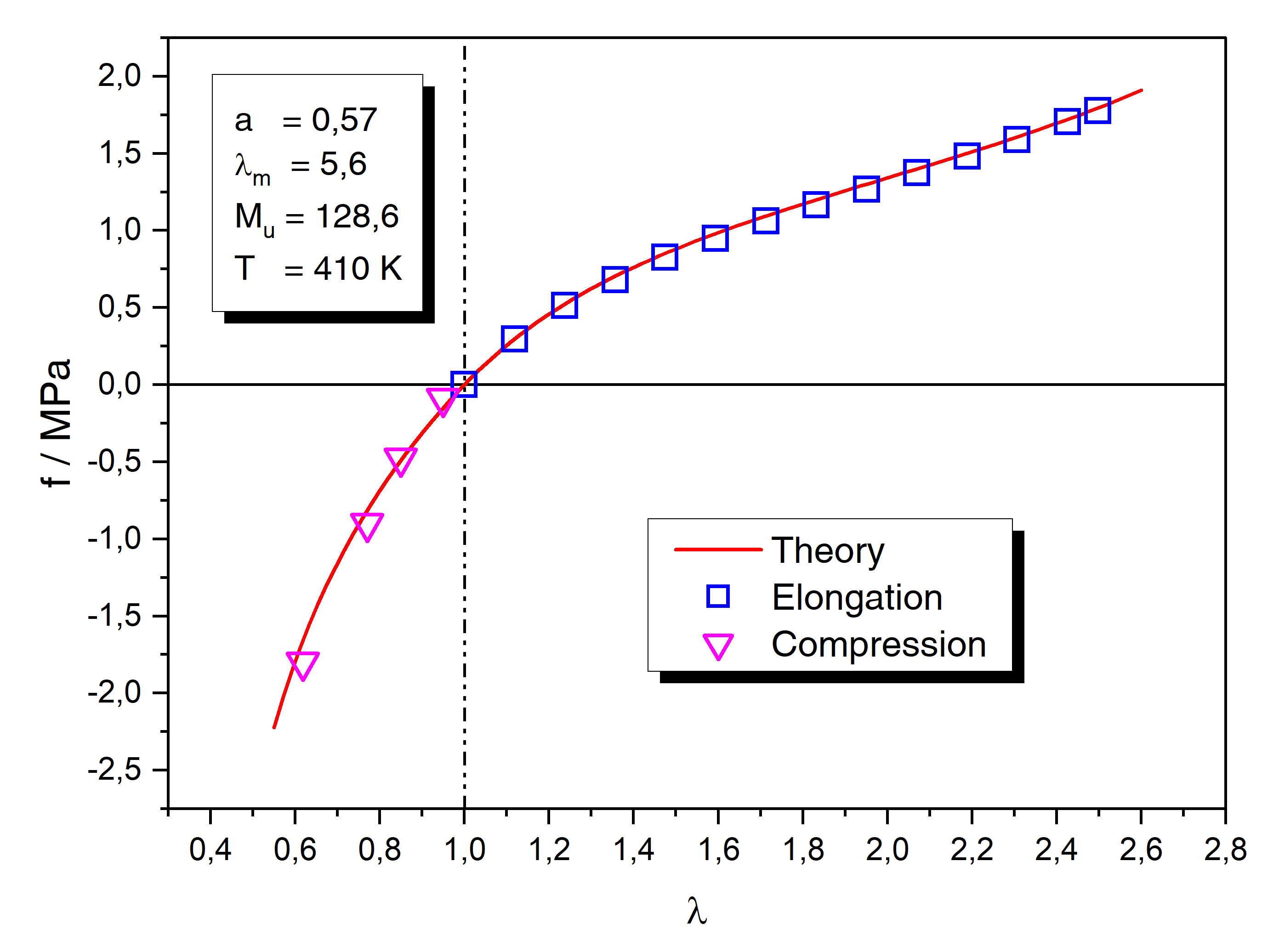} 
\caption{A quasi-static stress-strain curve of PMMA (polymethyl methacrylate), with stress (f) plotted in MPa versus strain $\lambda$ under elongation and compression with  parameters as indicated. The calculation (red line) is based on the van der Waals strain energy function eqn.(\ref{vdw1}) in uniaxial mode for elongation and biaxial mode for compression.}
\label{fig.1}
\end{figure}
\subsection{Irreversible Thermodynamics}
The Gibbs free energy density in equilibrium thermodynamics is written as
\begin{eqnarray}\label{gibbs}
dg &=& - sdT + fd\lambda
\end{eqnarray}
for constant pressure.
This Gibbs free energy density is generalized to account for time-dependent processes by introducing hidden variables that depend on both time and temperature
\begin{eqnarray}\label{ir1}
dg &=& - sdT + fd\lambda - \sum\limits_i A_i d\xi_i\quad .
\end{eqnarray}
We use the linear Onsager ansatz
\begin{eqnarray}\label{ir2}
\dot{\xi_i} &=& \alpha_{i} A_i
\end{eqnarray}
to describe the distance from thermodynamic equilibrium 
\cite{onsager_reciprocal_1931,onsager_reciprocal_1931-1,meixner_thermodynamische_1954}. 
The ${\bf A}_i$ are the generalized forces or affinities, the $\alpha_i$ are material dependent coefficients and the ${\dot{\xi}}_i$ the generalized
fluxes \cite{degroot_non-equilibrium_1962}. 
An ansatz of the Gibbs free energy density for isothermal and isobaric 
conditions may be the 
homogeneous quadratic form \cite{enderle_finite_1988}
  \begin{eqnarray}\label{ir1.1}
     g &=& \frac{1}{2}\,f_1\,w(\lambda)+ \sqrt{w(\lambda)}\,
     \sum\limits_i f^{(i)}_{12}\,\xi_i
         + \frac{1}{2}\,\sum\limits_i\,f^{(i)}_2\,\xi^2_i .
  \end{eqnarray}
With the help of the equations of state, the
mechanical equation of state
  \begin{eqnarray}\label{ir3}
   f & = &\left( \frac{\partial g}{\partial \lambda}\right )_{T,\xi_i}\quad ,
  \end{eqnarray}
the caloric equation of state
  \begin{eqnarray}\label{ir4}
   s & = & - \left( \frac{\partial g}{\partial T} 
   \right )_{{\lambda},\xi_i}
   \quad ,
  \end{eqnarray}
the internal equation of state
  \begin{eqnarray}\label{ir5}
   A_i & = & -\left( \frac{\partial g}{\partial \xi_i} 
   \right)_{{\lambda},T}
  \end{eqnarray}
and (\ref{ir2}), the hidden variables $\xi_i$ may be eliminated in (\ref{ir3}).
This yields a relationship for the nominal force $f$
\begin{eqnarray}\label{3}
     f(t)&=&G\cdot w'(t)\cdot \left\{ 1+\frac{\Gamma}{G}\cdot \left(1-M(t)\right )\right\}
\end{eqnarray}
with
\begin{eqnarray}
     M(t)&=&
     \frac{2}{\Gamma}
     \int \limits_{0}^{t} m(t-t') 
     {\left( \frac{w(t')}{w(t)}\right)}^{\frac{1}{2}}dt'
    \nonumber
\end{eqnarray}
where $\Gamma=G_g-G$ is the relaxation strength,
$G_g$ the maximum modulus, attained at highest frequencies 
$(G_g=G(\omega\to\infty))$ and $w'(t)=\partial_{\lambda} w(\lambda(t))$.
The expression $m(t-t')$ is the normalized relaxation--time spectrum
  \begin{eqnarray}
     m(t-t')&=&\sum_i \frac{h_i}{\tau_i}
     \,e^{\displaystyle -\frac{(t - t')}{\tau_i}}\quad .
  \end{eqnarray}
We want to focus on the comparison of the theory with experimental data, the theoretical calculations for the time dependent stress-strain curve are extensively represented in a recent paper \cite{ambacher_relaxation_1989}. 
The temperature dependence of the relaxation times obeys the WLF equation 
\cite{williams_temperature_1955} with constants e.g. for PMMA $C_1=6.34$, $C_2=63.62$ and 
$T_S=416.5$ K
  \begin{eqnarray}\label{3.1}
     \log a_T(T)&=&\frac{-C_1\cdot(T-T_S)}{C_2+T-T_S}\quad.
  \end{eqnarray}
\begin{figure}[ht]
\includegraphics[scale=0.37]{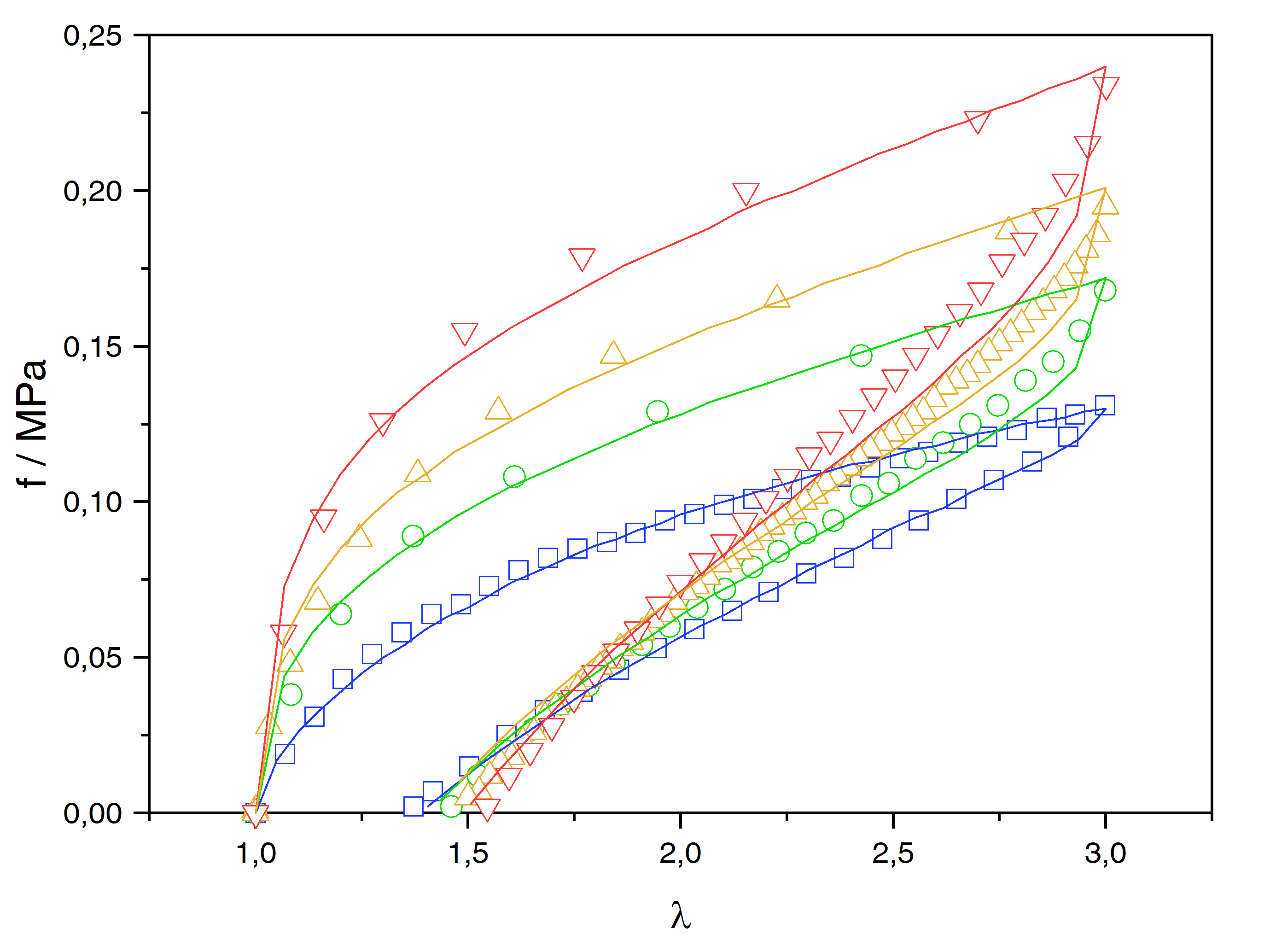}
\caption{Measurements of stress-strain curves of PVAc under elongation with different strain rates $\dot\epsilon$ ($\triangledown: 0,236$ 1/s, $\Delta: 0,120$ 1/s, $\circ: 0,06$ 1/s, $\square: 0,001$ 1/s) at a temperature of 333 K). Solid lines with the same color represent theory.}
\label{fig.2}
\end{figure}
\begin{figure}[ht]
\includegraphics[scale=0.38]{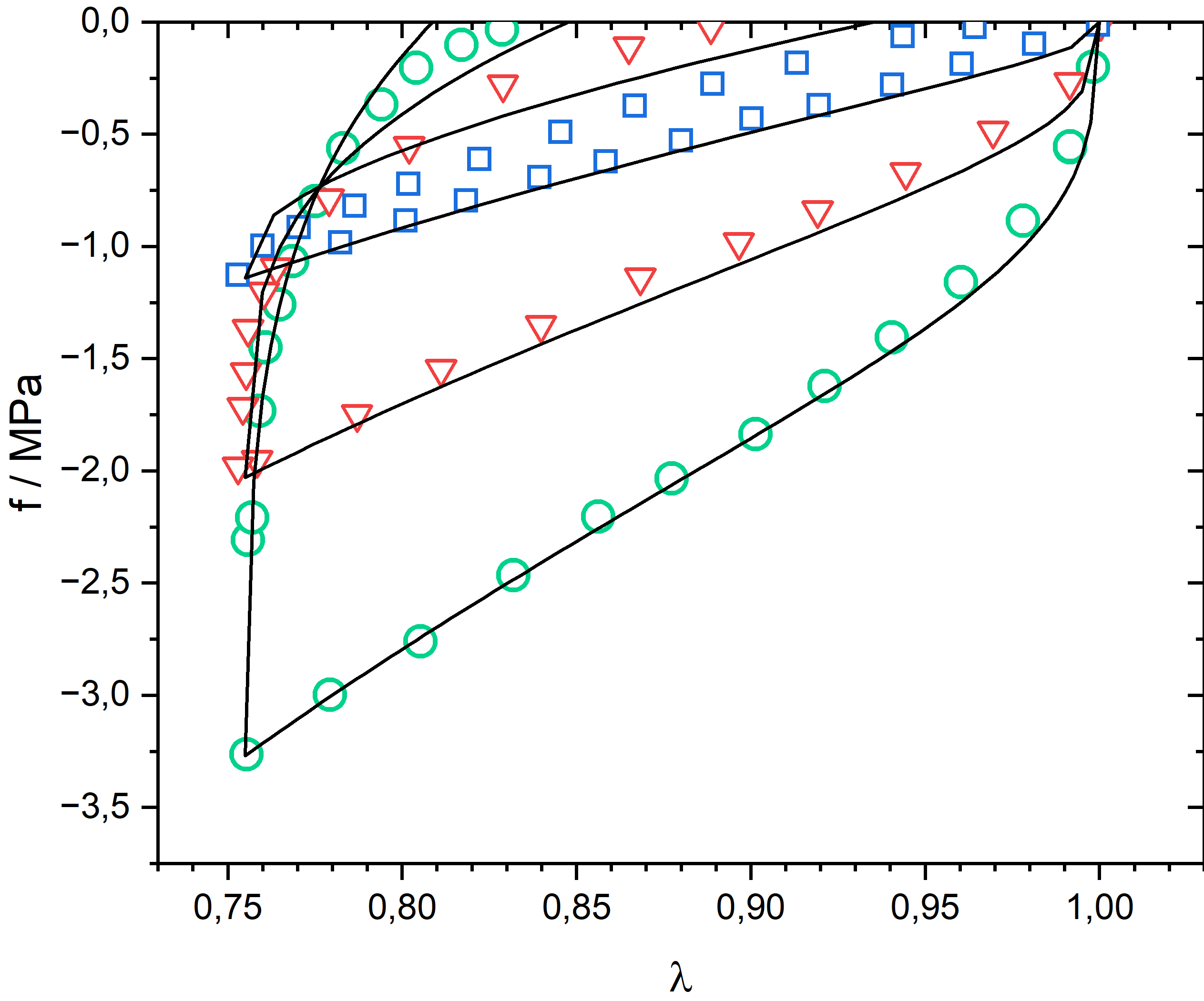}
\caption{Stress-strain curves of PMMA under biaxial compression at different strain rates $\dot \epsilon$ ($\square: 0,002$ 1/s, $\triangledown: 0,009$ 1/s $\circ: 0,035$ 1/s) at a temperature of $T=403 K$. Solid lines represent theory.
}
\label{fig.3}
\end{figure}
\section{Discussion}
Experimental results of the stress-strain curves are described by this theory (eqn.(\ref{vdw1}) and eqn.(\ref{3})) under isothermal conditions and above the glass transition (with relaxation time spectra of PVAc and PMMA respectively, measured at small strains).
As an example, the uniaxial stress-strain behavior of PVAc, shown in Figure \ref{fig.2}, demonstrates good agreement between the theoretical predictions and the experimental data. Similarly, Figure \ref{fig.3} presents the biaxial stress-strain response of PMMA, where the model also captures the main features of the observed behavior. In both cases, the theory reflects the characteristic deformation patterns of polymers in the rubbery regime, indicating its applicability to different loading modes under different thermal conditions and strain rates.
\section{Conclusions}
The phenomenological van der Waals approach
yields good 
accordance with the quasi static experiment over the whole range of strains
and deals with a small set of parameters, which may be interpreted 
on a molecular basis.\\
The formalism of irreversible thermodynamics leads us, independently
from any consideration about equilibrium stress strain curves,
straightforward to a relationship, which describes time and 
temperature dependent stress strain curves in a temperature range 
above the glass transition temperature without any residual fitting parameters. 
In this temperature region, the time-temperature equivalence is determined by the WLF equation.\\
Our calculations of the stress-strain cycles, for both uniaxial elongation and biaxial compression of permanent networks, are based on a single, material-specific relaxation time spectrum determined at small strains, and they produce remarkably accurate results.
This leads us to conclude that the relaxation time spectrum remains unaffected by variations in strain and strain rate, implying thermorheological simplicity of the material.\\
As the experiments approach the glass transition temperature, the shape of the stress-strain curves 
under extension and compression changes markedly \cite{mark_glass_2007,ogura_rheological_2013,bending_measurement_2014, siviour_high_2016, abdel-wahab_temperature-dependent_2017, roth_polymer_2017,medvedev_multistep_2022,berthier_modern_2023}. 
This presents a significant challenge: to assess whether, and under what conditions, the framework of irreversible thermodynamics remains applicable, and to evaluate whether the underlying deformation mechanisms can be more thoroughly understood.
\section*{Acknowledgments}
We like to thank Professor O. Marti
for 
his kind support and for allowing us to publish under the courtesy of the Institute of Experimental Physics at the University of Ulm. We also thank Stefan Golfier (MPI Dresden) for fruitful discussions. The authors gratefully acknowledge the late Professor H. G. Kilian, under whose tenure as chair this research was largely conducted. We would also like to thank the RÖHM Company (Darmstadt) for supplying the PMMA networks.



\begin{thebibliography}{10}
\expandafter\ifx\csname url\endcsname\relax
  \def\url#1{\texttt{#1}}\fi
\expandafter\ifx\csname urlprefix\endcsname\relax\def\urlprefix{URL }\fi
\expandafter\ifx\csname href\endcsname\relax
  \def\href#1#2{#2} \def\path#1{#1}\fi

\bibitem{flory_principles_1953}
P.~J. Flory, Principles of {Polymer} {Chemistry}, Cornell University Press, 1953.

\bibitem{treloar_physics_1967}
L.~R.~G. Treloar, The {Physics} of {Rubber} {Elasticity}, 2nd {Ed}., Clarendon Press, Oxford, 1967.

\bibitem{he_comparative_2022}
H.~He, Q.~Zhang, Y.~Zhang, J.~Chen, L.~Zhang, F.~Li, A comparative study of 85 hyperelastic constitutive models for both unfilled rubber and highly filled rubber nanocomposite material, Nano Materials Science 4~(2) (2022) 64--82.

\bibitem{vilgis_equilibrium_1986}
T.~Vilgis, H.~G. Kilian, Equilibrium conformations of polymer chains {Part} {II}. {Conformons} in networks, Colloid \& Polymer Science 264~(2) (1986) 137--142.

\bibitem{zrinyi_decisive_1989}
M.~Zrinyi, H.~G. Kilian, E.~Horkay, On the decisive role of finite chain extensibility and global interactions in networks, Colloid \& Polymer Science 267~(4) (1989) 311--322.

\bibitem{andrikakou_behaviour_2016}
P.~Andrikakou, K.~Vickraman, H.~Arora, On the behaviour of lung tissue under tension and compression, Scientific Reports 6~(1) (2016) 36642.

\bibitem{naumann_mechanical_2022}
J.~Naumann, N.~Koppe, U.~H. Thome, M.~Laube, M.~Zink, Mechanical properties of the premature lung: {From} tissue deformation under load to mechanosensitivity of alveolar cells, Frontiers in Bioengineering and Biotechnology 10 (2022) 964318.

\bibitem{ambacher_relaxation_1989}
H.~Ambacher, H.~F. Enderle, H.~G. Kilian, A.~Sauter, Relaxation in permanent networks, in: M.~Pietralla, W.~Pechhold (Eds.), Relaxation in {Polymers}, Progress in {Colloid} \& {Polymer} {Science}, Steinkopff, Darmstadt, 1989, pp. 209--220.

\bibitem{enderle_irreversible_1984}
H.~F. Enderle, H.~G. Kilian, T.~Vilgis, Irreversible deformation of macromolecular networks, Colloid and Polymer Science 262~(9) (1984) 696--704.

\bibitem{onsager_reciprocal_1931}
L.~Onsager, Reciprocal {Relations} in {Irreversible} {Processes}. {I}., Physical Review 37~(4) (1931) 405--426, publisher: American Physical Society.

\bibitem{onsager_reciprocal_1931-1}
L.~Onsager, Reciprocal {Relations} in {Irreversible} {Processes}. {II}., Physical Review 38~(12) (1931) 2265--2279, publisher: American Physical Society.

\bibitem{degroot_non-equilibrium_1962}
S.~R. deGroot, P.~Mazur, Non-equilibrium {Thermodynamics}, North Holland, Amsterdam, 1962.

\bibitem{lebon_understanding_2008}
G.~Lebon, D.~Jou, J.~Casas-Vázquez, Understanding non-equilibrium thermodynamics: foundations, applications, frontiers, Springer, Berlin, 2008.

\bibitem{bikkin_non-equilibrium_2021}
H.~Bikkin, I.~I. Lyapilin, Non-equilibrium {Thermodynamics} and {Physical} {Kinetics}, De Gruyter, 2021, publication Title: Non-equilibrium Thermodynamics and Physical Kinetics.

\bibitem{di_vita_non-equilibrium_2022}
A.~Di~Vita, Non-equilibrium {Thermodynamics}, Vol. 1007 of Lecture {Notes} in {Physics}, Springer International Publishing, Cham, 2022.

\bibitem{kraus_relaxation_1994}
V.~Kraus, H.~G. Kilian, M.~Saile, Relaxation mode coupling and universality in stress-strain cycles of networks including the glass transition region, Polymer 35~(11) (1994) 2348--2354.

\bibitem{kraus_relaxation_1992}
V.~Kraus, H.~G. Kilian, W.~v. Soden, Relaxation in permanent networks, in: S.~Wartewig, G.~Helmis (Eds.), Physics of {Polymer} {Networks}, Progress in {Colloid} \& {Polymer} {Science}, Steinkopff, Darmstadt, 1992, pp. 27--36.

\bibitem{wrana_introduction_2009}
C.~Wrana, Introduction to {Polymer} {Physics}: 100 {Years} {Synthetic} {Rubber}, {Creating} the {Way} the {World} {Moves} {Today}, Lanxess AG, 2009.

\bibitem{kilian_thermomechanik_1979}
H.-G. Kilian, Thermomechanik molekularer {Netzwerke}, Physikalische Blätter 12 (1979) 642.

\bibitem{kilian_molecular_1980}
H.~G. Kilian, A molecular interpretation of the parameters of the van der {Waals} equation of state for real networks, Polymer Bulletin 3~(3) (1980) 151--158.

\bibitem{kilian_equation_1981}
H.-G. Kilian, Equation of state of real networks, Polym 22 (1981) 209.

\bibitem{eisele_new_1981}
U.~Eisele, B.~Heise, H.-G. Kilian, M.~Pietralla, A {New} {Method} of {Characterizing} {Molecular} {Networks} with the van der {Waals} {Equation} of {State}, Angewandte Makromolekulare Chemie 100~(1) (1981) 67--85.

\bibitem{kilian_rubber_1986}
H.~G. Kilian, K.~Unseld, Rubber elasticity and network structure, Colloid and Polymer Science 264~(1) (1986) 9--18.

\bibitem{hamm_uniaxiale_1993}
W.~A. Hamm, Uniaxiale {Kompression} gefüllter {Vulkanisate}, {PhD} {Thesis}, Universität Ulm (1993).

\bibitem{meixner_thermodynamische_1954}
J.~Meixner, Thermodynamische {Theorie} der elastischen {Relaxation}, Zeitschrift für Naturforschung A 9a (1954) 654.

\bibitem{enderle_finite_1988}
H.-F. Enderle, Finite {Viskoelastizität} von {Van} der {Waals}– {Netzwerken}, {PhD} {Thesis}, Universität Ulm (1988).

\bibitem{williams_temperature_1955}
M.~L. Williams, R.~F. Landel, J.~D. Ferry, The {Temperature} {Dependence} of {Relaxation} {Mechanisms} in {Amorphous} {Polymers} and {Other} {Glass}-forming {Liquids}, Journal of the American Chemical Society 77 (1955) 3701.

\bibitem{mark_glass_2007}
D.~J. Plazek, K.~L. Ngai, The {Glass} {Temperature}, in: J.~E. Mark (Ed.), Physical {Properties} of {Polymers} {Handbook}, Springer New York, New York, NY, 2007, pp. 187--215.

\bibitem{ogura_rheological_2013}
K.~Ogura, M.~H. Wagner, Rheological characterization of cross-linked poly(methyl methacrylate), Rheologica Acta 52~(8-9) (2013) 753--765.

\bibitem{bending_measurement_2014}
B.~Bending, K.~Christison, J.~Ricci, M.~D. Ediger, Measurement of {Segmental} {Mobility} during {Constant} {Strain} {Rate} {Deformation} of a {Poly}(methyl methacrylate) {Glass}, Macromolecules 47~(2) (2014) 800--806, publisher: American Chemical Society.

\bibitem{siviour_high_2016}
C.~R. Siviour, J.~L. Jordan, High {Strain} {Rate} {Mechanics} of {Polymers}: {A} {Review}, Journal of Dynamic Behavior of Materials 2~(1) (2016) 15--32.

\bibitem{abdel-wahab_temperature-dependent_2017}
A.~A. Abdel-Wahab, S.~Ataya, V.~Silberschmidt, Temperature-dependent mechanical behavior of {PMMA}: {Experimental} analysis and modeling, Polymer Testing 58 (2017) 86--95, publisher: Loughborough University.

\bibitem{roth_polymer_2017}
C.~B. Roth (Ed.), Polymer glasses, CRC Press, Taylor \& Francis Group, Boca Raton, 2016.

\bibitem{medvedev_multistep_2022}
G.~A. Medvedev, E.~Xing, M.~D. Ediger, J.~M. Caruthers, Multistep {Deformation} {Experiment} and {Development} of a {Model} for the {Mechanical} {Behavior} of {Polymeric} {Glasses}, Macromolecules 55~(15) (2022) 6351--6363, publisher: American Chemical Society.

\bibitem{berthier_modern_2023}
L.~Berthier, D.~R. Reichman, Modern computational studies of the glass transition, Nature Reviews Physics (Jan. 2023).

\end{thebibliography}
\end{document}